\pdfoutput=1 
\documentclass[twocolumn]{JINST}
\usepackage{graphicx}
\usepackage{subfigure}
\usepackage{float}

\newcommand{\refeq}[1]{Eq. (\ref{#1})}

\newcommand{\reffig}[1]{Figure \ref{#1}}

\newcommand{\ATS}{A_{TS}}
\newcommand{\ASYN}{A_{SYN}}
\newcommand{\VTS}{V_{TS}}
\newcommand{\VSYN}{V_{SYN}}
\newcommand{\calR}{\mathcal{R}}
\newcommand{\Teij}{T_e^{ij}}
\newcommand{\sigtij}{\sigma_{total}^{ij}}
\newcommand{\sigtmn}{\sigma_{total}^{mn}}

\newcommand{\SNRTe}{SNR_{T_e}}
\newcommand{\SNRtotali}{SNR_{total}^i}
\newcommand{\SNRbgi}{SNR_{bg}^i}

\title{Quantifying noise sources in the KSTAR 2014 Thomson Scattering system from the measured variation on electron temperature}

\author{Tae-suk~Oh$^{1, *}$, K.H.~Kim$^1$, J.H~LEE$^2$, S.H~LEE$^2$, R.~SCANNELL$^3$, A.~R.~Field$^3$, K.~Cho$^4$, M.~S.~BAWA'ANEH$^{5,~6}$ and Y.-c.~GHIM$^{1, **}$\\
\llap{$^1$}Department of Nuclear and Quantum Engineering, KAIST, Daejeon, Korea\\
\llap{$^2$}National Fusion Research Institute, Daejeon, Korea\\
\llap{$^3$}Culham Centre for Fusion Energy, Culham Science Centre Abingdon, Oxfordshire OX143DB, U.K.\\
\llap{$^4$}Department of Physics, Sogang University, Seoul, Korea\\
\llap{$^5$}Department of Physics, The Hashemite University, Zarqa, Jordan\\
\llap{$^6$}Department of Applied Mathematics and Sciences, Khalifa University, UAE\\

E-mail: \email{$^*$taeseokoh@kaist.ac.kr, $^{**}$ycghim@kaist.ac.kr}}

\abstract{With the Thomson scattering (TS) system in KSTAR, temporal evolution of electron temperature ($T_e$) is estimated using a  weighted look-up table method with fast sampling ($1.25$ or $2.5$ GS/s) digitizers during the 2014 KSTAR campaign. Background noise level is used as a weighting parameter without considering the photon noise due to the absence of information on absolute photon counts detected by the TS system. Estimated electron temperature during a relatively quiescent discharge are scattered, i.e., $15$\% variation on $T_e$ with respect to its mean value. We find that this $15$\% variation on $T_e$ cannot be explained solely by the background noise level which leads us to include photon noise effects in our analysis. Using synthetic data, we have estimated the required photon noise level consistent with the observation and determined the dominant noise source in KSTAR TS system.}

\keywords{Thomson scattering, variation in electron temperature, dominant noise source, photon counts}

\begin{document}
\section{Introduction}\label{sec:intro}
\begin{figure}[!]
\centering
\includegraphics[width = .45\textwidth, height = 7cm]{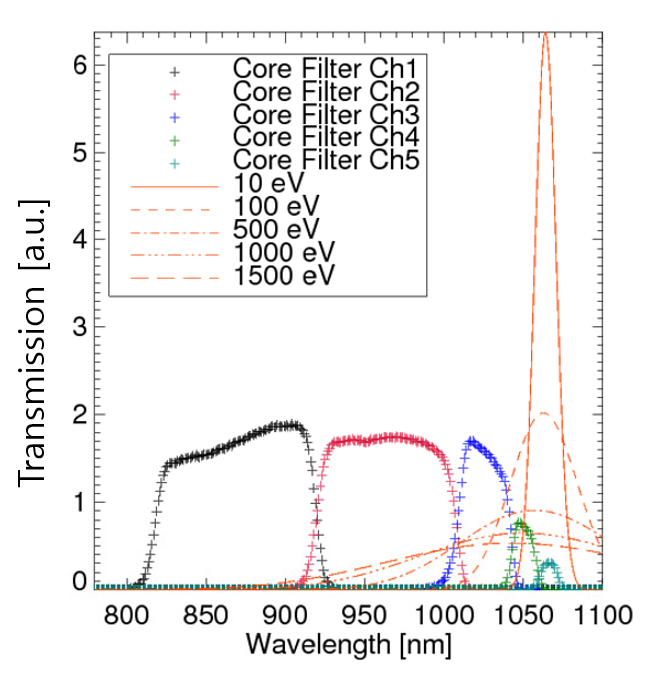}
\caption{Measured filter functions (`+'), i.e., relative transmittance coefficients of the optical band-pass filters as a function of the wavelength, for five separate channels of a core polychromator. Channel 1 looks at the shortest wavelength range, while channel 5 contains the Nd-YAG laser wavelength of $1064$ nm. Also shown (lines) are the spectral distributions of the scattered photons \cite{O_Naito_1993} for various temperatures.}
\label{fig:KSTAR_filter_fcn}
\end{figure}

Thomson scattering (TS) systems are widely used to measure the temperature ($T_e$) and density ($n_e$) of electrons in fusion devices. KSTAR Thomson scattering system consists of separate core and edge collection optics collecting the Thomson scattered photons which are transmitted to polychromators via optical fibres \cite{Oh_RSI_2012,Lee_RSI_2010}. After the band-pass filters, where \reffig{fig:KSTAR_filter_fcn} shows an example of the measured filter functions, optical signals are converted to electrical signals and recorded with the newly installed fast sampling digitizers (NI PXIe-5160) with the sampling rate of either $1.25$ or $2.5$ GS/s, in addition to the existing charge-integrating (gating) digitizer.

We use a weighted look-up table method to estimate the time evolution of electron temperature while exclusively considering the uncertainty due to the background noise. Another possibly large source of uncertainty may originate from the photon (Poisson) noise. However, the photon noise could not be reflected in our analysis due to the lack of absolute photon counts. Note that the absolute photon counts can be estimated based on Rayleigh or rotational Raman calibration \cite{LeBlanc_RSI_2008,Scannell_Thesis}, but such data have not been obtained with the fast sampling digitizers during the 2014 KSTAR campaign. Rayleigh calibration data are obtained only with the existing charge-integrating (gating) digitizers.

Estimated electron temperature shows scatters, i.e., $15$\% variation on $T_e$ with respect to its mean value. By using the synthetic data, we have found that this scattering on $T_e$ cannot be explained solely by the uncertainty due to the background noise. Our main goal of this paper is to estimate the `total' noise level of the Thomson signal consistent with the observed variation on $T_e$ using the synthetic data. With the levels of estimated total noise and the measured background noise, we deduce the photon noise level assuming that photon noise and background noise are uncorrelated. We confirm that photon noise prevails over the background one for the Thomson data from the 2014 KSTAR campaign.
\section{Estimating electron temperature}\label{sec:estimating_T_e}
A polychromator contains five channels consisting of five band-pass filters shown in \reffig{fig:KSTAR_filter_fcn} with five photon detectors (IR-enhanced Si APD: Hamamatsu S11519).
\begin{figure}[t]
\centering
\subfigure[]
{
\includegraphics[width = .3\textwidth, height = 4.9cm]{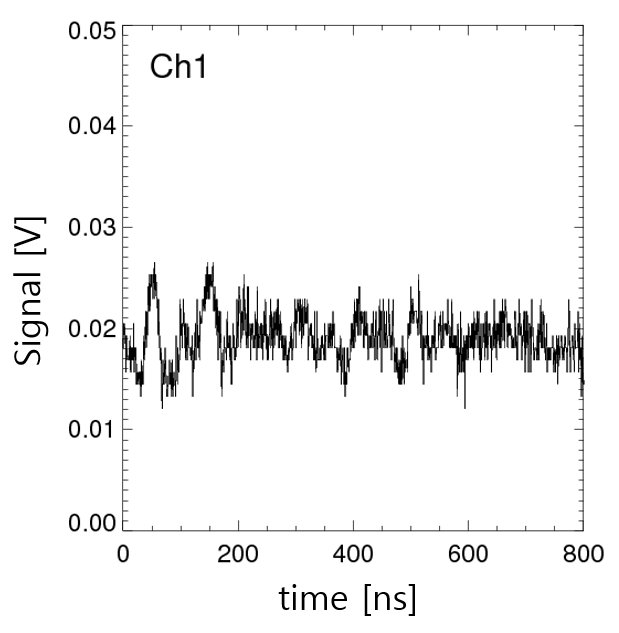}\label{fig:signal_ch1}
}
\subfigure[]
{
\includegraphics[width = .3\textwidth, height = 5cm]{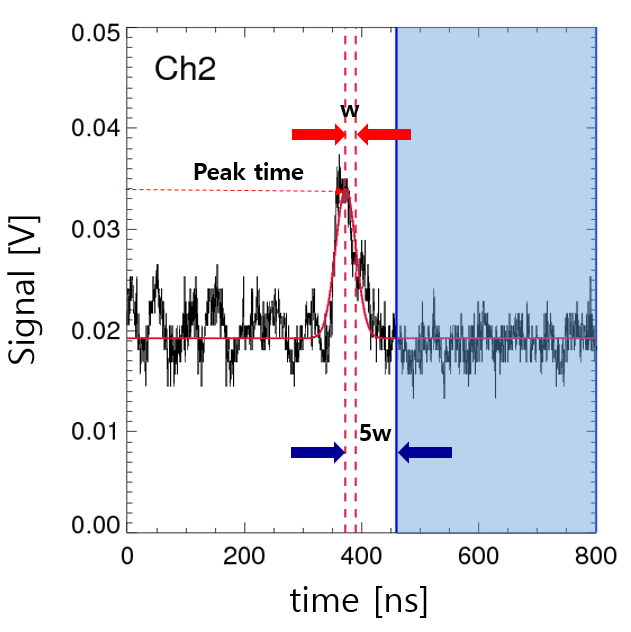}\label{fig:signal_ch2}
}
\subfigure[]
{
\includegraphics[width = .3\textwidth, height = 5cm]{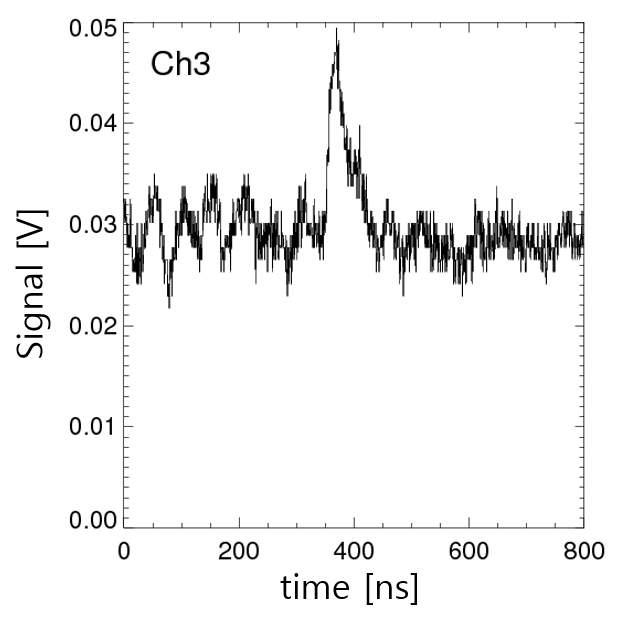}\label{fig:signal_ch3}
}
\\
\subfigure[]
{
\includegraphics[width = .3\textwidth, height = 5cm]{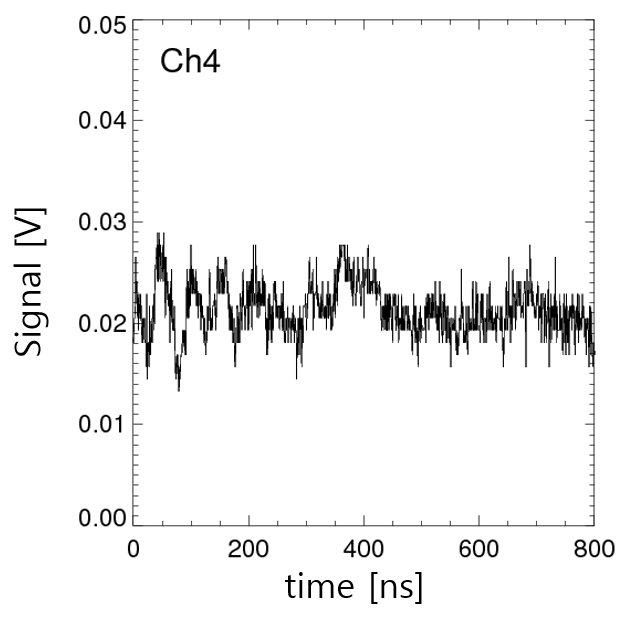}\label{fig:signal_ch4}
}
\subfigure[]
{
\includegraphics[width = .3\textwidth, height = 5cm]{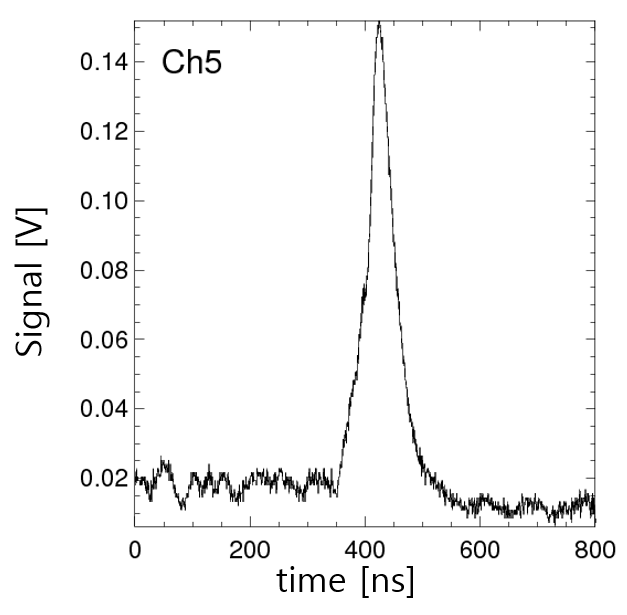}\label{fig:signal_ch5}
}
\caption{Measured Thomson signal from each channel for KSTAR shot \#10433 at plasma time = 10 s. Thomson signal from ch 1 cannot be distinguished from the noise, while the signal from ch 5 is dominated by the straylight. Red line in (b) illustrates the Gaussian fitting with the width of $w$ to the raw Thomson signal (black). Blue shaded region is used to estimate the background noise level of ch 2.}
\label{fig:signal}
\end{figure}
\reffig{fig:signal} shows an example of the Thomson signal from each channel. Although the background noise levels are quite similar for all five channels, ch 1 does not show observable Thomson scattered signal. This may be due to electron temperature being less than $1$ keV (see \reffig{fig:KSTAR_filter_fcn}). The signal from ch 5 is mostly due to the straylight as attested by the fact that the signal levels before and during plasma shots are identical (not shown in the figure). The level of Thomson signal from ch 4 is smaller compared to signals obtained from chs 2 and 3, again perhaps due to not-high enough electron temperature.

To estimate the integral of the $i$th channel Thomson signal $\ATS^i$, we perform a Gaussian fitting to the Thomson signal $\VTS^i$ (red line in \reffig{fig:signal_ch2}) which can be written as \cite{Scannell_Thesis} 
\begin{equation} \label{eq:V_ts}
\VTS^i\left( t\right)= G \:n_{e} \:n_{laser}\left( t\right)\: \frac{\mathrm{d} \sigma _{TS}}{\mathrm{d} \Omega }\:\Delta Q\:L\:T(\lambda _{L})\:QE\:\int d\lambda \:\frac{\phi^i (\lambda )}{\phi (\lambda _L)}\: \frac{S(\lambda ,T_e,\theta )}{\lambda _L},
\end{equation}
where $G$ is the APD gain factor, $n_{laser}\left( t\right)$ the number of photons per unit time as a function of time where $\int dt\:n_{laser}\left( t\right)=N_{laser}$ is the total number of photons in a single laser pulse. $\frac{\mathrm{d} \sigma _{TS}}{\mathrm{d} \Omega }$ is the differential Thomson scattering cross-section area, $\Delta Q$ the solid angle of the TS system and $L$ the scattering length. The net transmission coefficient of the collective optics $T(\lambda)$ is a function of the wavelength, but we assume that it is constant within the range of interests; hence $T(\lambda)=T(\lambda_L)$ where $\lambda_L$ is the laser wavelength. As the $i$th channel filter function $\phi^i(\lambda)$ includes the wavelength variation of the quantum efficiency $QE$ of the APD detector, the value of $QE$ is taken at the laser wavelength. $S(\lambda ,T_e,\theta )$ is the spectral distribution of the scattered photons \cite{O_Naito_1993} where $\theta$ is the scattering angle. Then, $\ATS^i$ is simply
\begin{equation} \label{eq:A_ts}
\ATS^i= \int dt \VTS^i\left( t\right).
\end{equation}

\begin{figure}[t]
\centering
\subfigure[]
{
	\includegraphics[width = .45\textwidth,height=6.4cm]{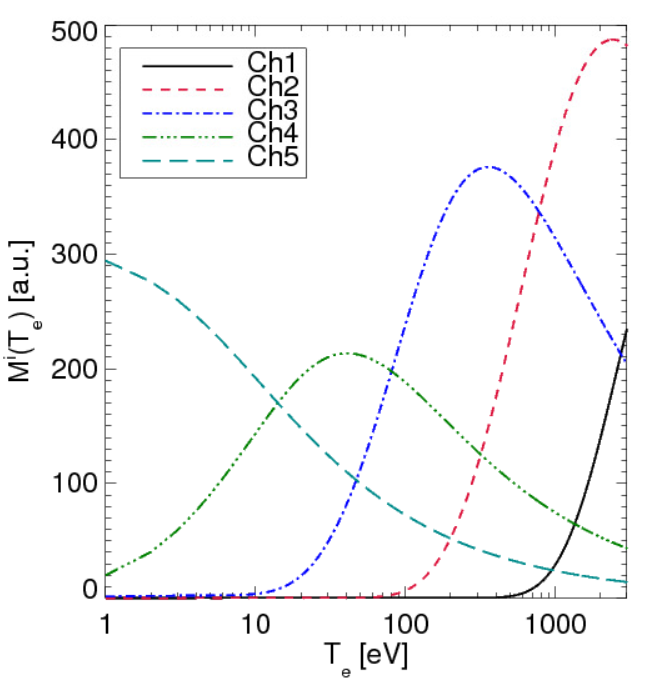}\label{fig:lookup_a}
}
\subfigure[]
{
	\includegraphics[width = .45\textwidth,height=6.4cm]{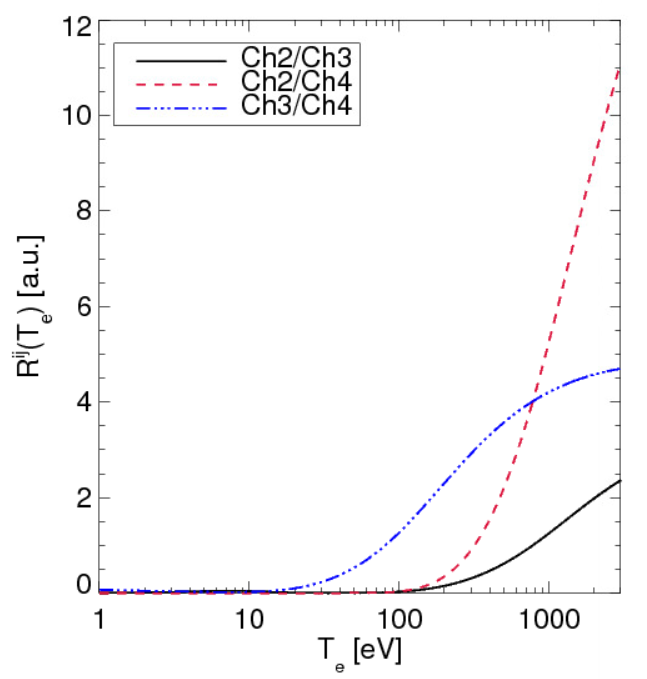}\label{fig:lookup_b}
}
\caption{(a) $M^i\left( T_e\right)$ and (b) $\calR^{ij}\left( T_e\right)$ the ratio of $\ATS^i$ between channels of 2, 3 and 4.}
\label{fig:lookup}
\end{figure}

From \refeq{eq:V_ts}, we find that the terms outside the integral do not depend on the channels. Thus, by taking the ratio of measured signals between the two channels such as
\begin{equation}
\label{eq:ts_ratio}
\frac{\ATS^i}{\ATS^j}=\frac{\int d\lambda \:\phi^i (\lambda )\: S(\lambda ,T_e,\theta )}{\int d\lambda \:\phi^j (\lambda )\: S(\lambda ,T_e,\theta )}\equiv\calR^{ij}\left( T_e, \theta \right),
\end{equation}
we can construct an equation that depends on only $T_e$ and $\theta$. Note that we have $\theta\approx90^\circ$ in this study \cite{Oh_MST_2011}. Defining $M^i\left( T_e\right)=\int d\lambda \:\phi^i (\lambda )\: S(\lambda ,T_e,\theta=90^\circ)$, \reffig{fig:lookup_a} shows $M^i$ as a function of $T_e$ for all five channels. As we decide not to use signals from chs 1 and 5 due to indiscernible Thomson signals from the background noise or stray light, we create $\calR^{ij}$ with chs 2, 3 and 4 which are shown in \reffig{fig:lookup_b}.

\begin{figure}[t]
\centering
\subfigure[]
{
	\includegraphics[width = .42\textwidth,height=6.5cm]{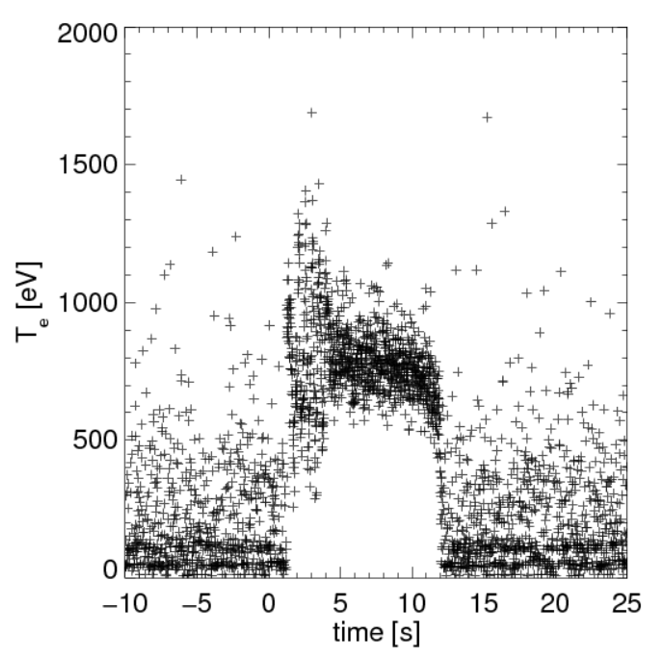}
}    
\subfigure[]
{
	\includegraphics[width = .42\textwidth,height=6.5cm]{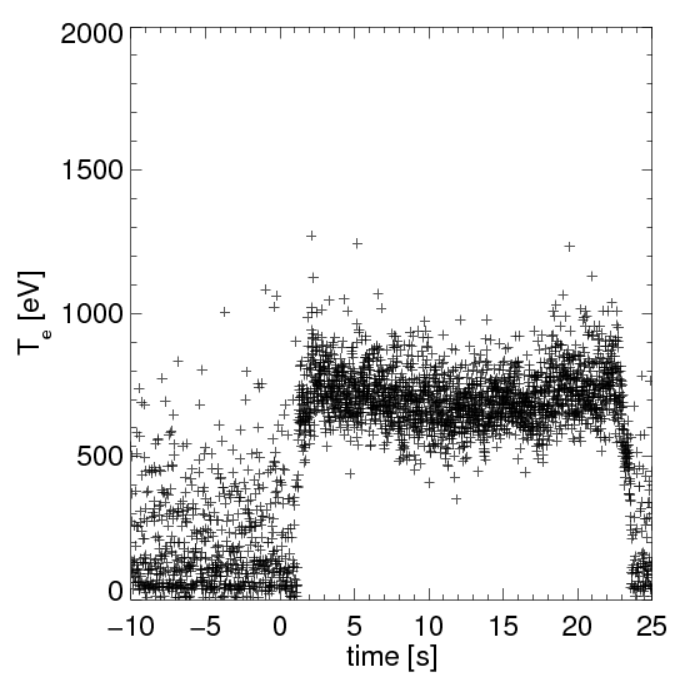}
}  
\caption{Examples of estimated $T_e$ using a weighted look-up table method for (a) KSTAR shot \#10401 and (b) \#10433.}
\label{fig:T_e}
\end{figure}

Because we may find three different $T_e$'s based on the measurements from \reffig{fig:lookup_b}, the $T_e$'s can be averaged with a weighting factor of the inverse of corresponding total uncertainty as
\begin{equation} \label{eq:electron_temperature}
T_e = \sum_{i,j = 2}^{4} \Teij\frac{1}{\sigtij}\left(\sum_{m,n = 2}^{4}\frac{1}{\sigtmn}\right)^{-1},\quad \quad (j>i,\:n>m),
\end{equation}
where $\Teij$ and $\sigtij$ are the estimated electron temperature using $\calR^{ij}$ and its total uncertainty, respectively. $\sigtij$ can be estimated using a propagation of uncertainty assuming that the total uncertainty of the $i$th and $j$th channels, $\sigma_{total}^i$ and $\sigma_{total}^j$, are uncorrelated. Ideally, $\sigma_{total}^i$ needs to include both the background noise $\sigma_{bg}^i$ and the photon noise $\sigma_{ph}^i$:
\begin{equation}
\label{eq:total_noise}
\left( {\sigma_{total}^i}\right)^2=\left({\sigma_{bg}^i}\right)^2+\left({\sigma_{ph}^i}\right)^2. 
\end{equation}
However, not knowing the photon noise level, we estimate the $T_e$ by setting ${\sigma_{total}^i}=\sigma_{bg}^i$. In this paper, we estimate the background noise of the $i$th channel as 
\begin{equation}
\label{eq:bg_uncertainty}
\sigma_{bg}^i = \delta^i \sqrt{N_{TS}^i} \Delta t,
\end{equation}
where $\delta^i$ is the standard deviation of the data in the blue shaded region shown in \reffig{fig:signal_ch2}. As shown in \reffig{fig:signal_ch2}, we first fit a Gaussian function to the Thomson signal which finds the amplitude, peak-time and width ($w$) of the Gaussian. The blue shaded region where we obtain the background noise level starts at $5w$ away from the peak-time assuming that no Thomson signal exists in this region, hence only the background signal. $N_{TS}^i$ is the number of data points within the fitted Gaussian, i.e., $N_{TS}^i = 4w/\Delta t$, where $\Delta t$ is $0.8$ ns, the time step of the data points with $1.25$ GS/s. \reffig{fig:T_e} shows examples of the temporal evolution of $T_e$ for two different KSTAR plasma shots estimated using \refeq{eq:electron_temperature}.
\section{Synthetic Thomson data and signal-to-noise ratio}
As the estimated $T_e$ shows about $15$\% variation on $T_e$ with respect to its mean value (e.g. see \reffig{fig:T_e}), we raise the following questions: 1) can the background noise explain such an observed variation on $T_e$? 2) if not, how large does $\sigma_{ph}^i$ have to be to explain the observation? To be able to answer these questions, we generate synthetic Thomson signal where we can vary the noise levels numerically.

By letting $n_{laser}\left( t\right)$ in \refeq{eq:V_ts} have a Gaussian form in a time domain (consistent with fitting the Gaussian function to measured Thomson signals), we generate the synthetic Thomson data for each channel similar to the measured data shown in \reffig{fig:signal}. We set the time step of synthetic data to be $0.8$ ns corresponding to $1.25$ GS/s of the fast sampling digitizer. A random number selected from a normal distribution is, then, added to each time point as a noise. The width of this normal distribution $\delta_{SYN}$ is set such that  $\sigma_{SYN}=\delta_{SYN}\sqrt{N_{TS}^i}\Delta t$ (cf. \refeq{eq:bg_uncertainty}) controlling the noise level. This completes generating synthetic Thomson signal for a single laser pulse. Let us denote the $i$th channel synthetic Thomson signal as $\VSYN^i\left( t\right)$ and its integral as $\ASYN^i$ estimated by fitting a Gaussian function to the $\VSYN^i\left( t\right)$ as if it were real measured signal.

\begin{figure}[t]
\centering
\subfigure[]
{
	\includegraphics[width = .45\textwidth,height=6.5cm]{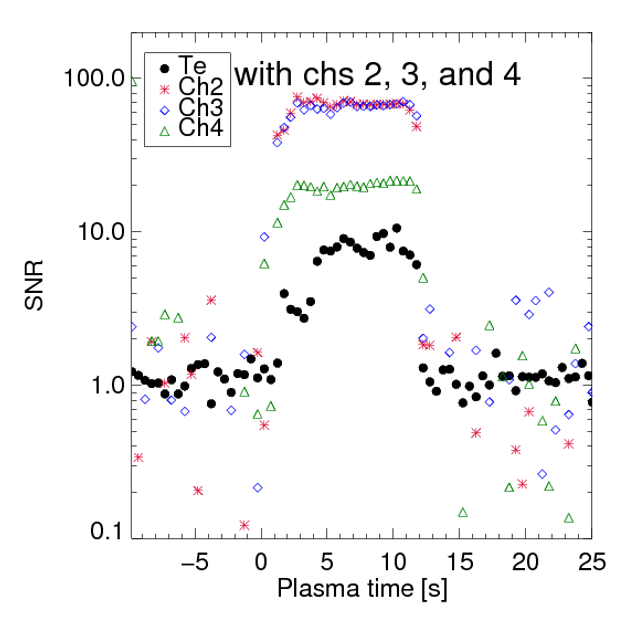}
}
\subfigure[]
{
	\includegraphics[width = .45\textwidth,height=6.5cm]{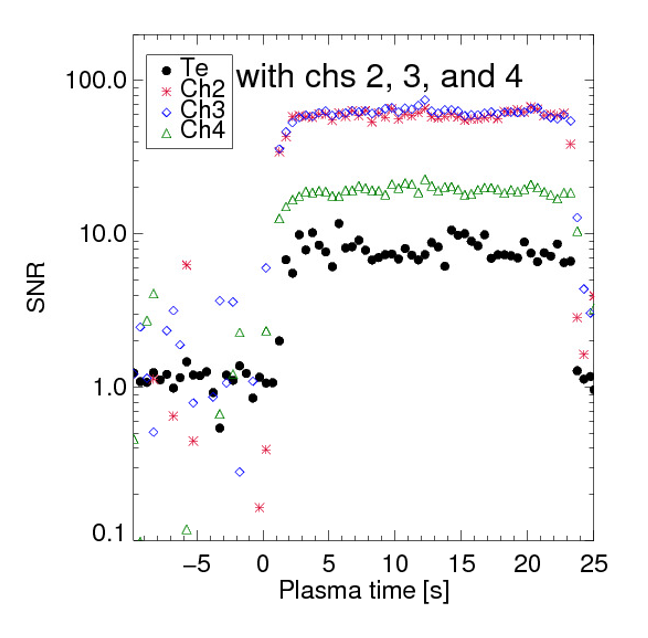}
}
\caption{Examples of $\SNRTe$ (black circle)  and $\SNRbgi$ for $i=2$ (red asterisk), $3$ (blue diamond) and $4$ (green triangle) for (a) KSTAR shot \#10401 and (b) \#10433. Here, $\left< T_e\right>$ is estimated using chs 2, 3, and 4.}
\label{fig:SNR_channel_T_e}
\end{figure}

We define the signal-to-noise ratio (SNR) of $T_e$, $\SNRTe$, and that of the measured (synthetic) Thomson signal from the $i$th channel, $SNR_*^i$ ($SNR_{SYN}^i$), to be
\begin{equation}
\label{eq:SNR}
\SNRTe=\frac{\left< T_e\right>}{\sigma_{T_e}},\qquad SNR_*^i = \frac{\left< \ATS^i \right>}{\left< \sigma_*^i\right>}, \qquad SNR_{SYN}^i = \frac{\left< \ASYN^i \right>}{\left< \sigma_{SYN}^i\right>},
\end{equation}
where the subscript $*$ takes any one of the total (`total'), background (`bg') or photon (`ph') for the noise source, i.e., $\SNRbgi$ indicates the SNR of the Thomson signal due to the background noise. Here, $\left< \cdot\right>$ indicates the time average with $50$ consecutive data points equivalent to the time duration of $0.5$ seconds, a couple of equilibrium evolution time scale for a typical KSTAR plasma, with the laser pulse repetition rate of $100$ Hz during the 2014 KSTAR campaign. $\sigma_{T_e}$ is the standard deviation of $T_e$ during the $0.5$ seconds. \reffig{fig:SNR_channel_T_e} shows time evolution of $\SNRTe$ and $\SNRbgi$ for $i=2$, $3$ and $4$ for the plasma shots shown in \reffig{fig:T_e}. 
\section{Estimating photon noise level and photon counts}
\begin{figure}[t]
\centering
\includegraphics[width = .5\textwidth,height=7cm]{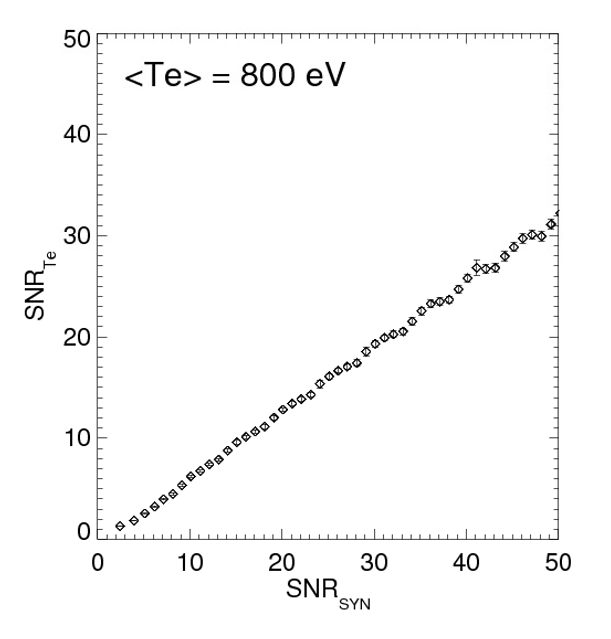}
\caption{With synthetic data of chs 2 and 3, $\SNRTe$ is estimated as a function of $SNR_{SYN}$ at $\left< T_e \right>=800$ eV. This allows us to obtain the required values of $SNR_{SYN}^{i}$ and the corresponding $\sigma_{SYN}^i$ given the experimental $\SNRTe$.}
\label{fig:synthetic_SNR_relation}
\end{figure}

We wish to find $SNR_{total}^i$ and the corresponding $\sigma_{total}^i$ for $i=2,3$ and $4$ consistent with the experimentally measured $\SNRTe$. Once we have $\sigma_{total}^i$, $\sigma_{ph}^i$ can be calculated using \refeq{eq:total_noise} with the measured $\sigma_{bg}^i$. Thus, we obtain $SNR_{total}^i$ and $\sigma_{total}^i$ with the synthetic data by recognizing that $SNR_{total}^i=SNR_{SYN}^i$ and $\sigma_{total}^i=\sigma_{SYN}^i$. Here, we face a problem: there may exist infinite number of solutions on the combination of $SNR_{total}^i$ for $i=2,3$ and $4$ consistent with the measured $\SNRTe$. To circumvent such a problem, anticipated by the experimental observation, we generate synthetic data only for chs 2 and 3 and let $SNR_{SYN}^{i=2}=SNR_{SYN}^{i=3}$ since $SNR_{bg}^4<SNR_{bg}^2\sim SNR_{bg}^3$ (see \reffig{fig:SNR_channel_T_e}). We generate a database of $\SNRTe=f\left(SNR_{SYN}^i, T_e\right)$ for many different values of $\left< T_e\right>$ for $i=2$ and $3$. \reffig{fig:synthetic_SNR_relation} shows the level of $\SNRTe$ estimated with the synthetic data as a function of $SNR_{SYN}^i$ at $\left< T_e \right>=800$ eV.

\begin{figure}[t]
\centering
\subfigure[]
{
\includegraphics[width = .43\textwidth, height = 6.5cm]{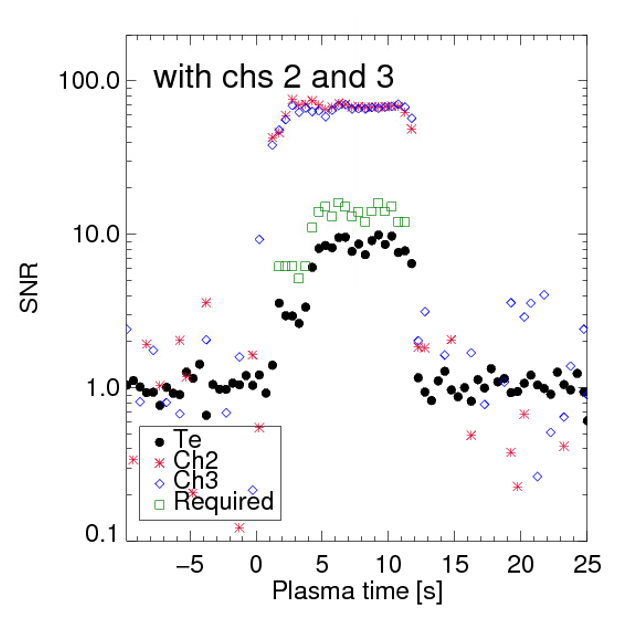}
}
\subfigure[]
{
\includegraphics[width = .43\textwidth, height = 6.5cm]{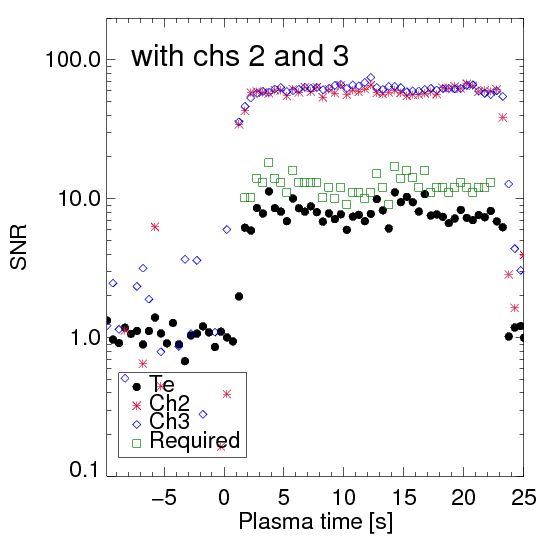}
}
\caption{Experimentally measured $\SNRTe$ (black circle) and $\SNRbgi$ for $i=2$ (red asterisk) and $3$ (blue diamond) for (a) KSTAR shot \#10401 and (b) \#10433. Green squares show the `required' $\SNRtotali$ estimated with the synthetic data consistent with the observed $\SNRTe$. Here, $\left< T_e\right>$ is estimated using chs 2 and 3 only.}
\label{fig:proper_snr}
\end{figure}

As the $\SNRTe$ from the synthetic data has been estimated with only chs 2 and 3 (e.g. \reffig{fig:synthetic_SNR_relation}), we recalculate the experimental $\SNRTe$ using only these two channels as shown in \reffig{fig:proper_snr}. Here, we also plot the `required' level of $SNR_{total}^i$ (green square) using the database of $\SNRTe=f\left(SNR_{SYN}^i, T_e\right)$ which allows us to estimate $\sigma_{total}^i$ for $i=2$ and $3$  using \refeq{eq:SNR}.

\begin{figure}[tp]
\centering
\subfigure[]
{
\includegraphics[width = .43\textwidth, height = 6.5cm]{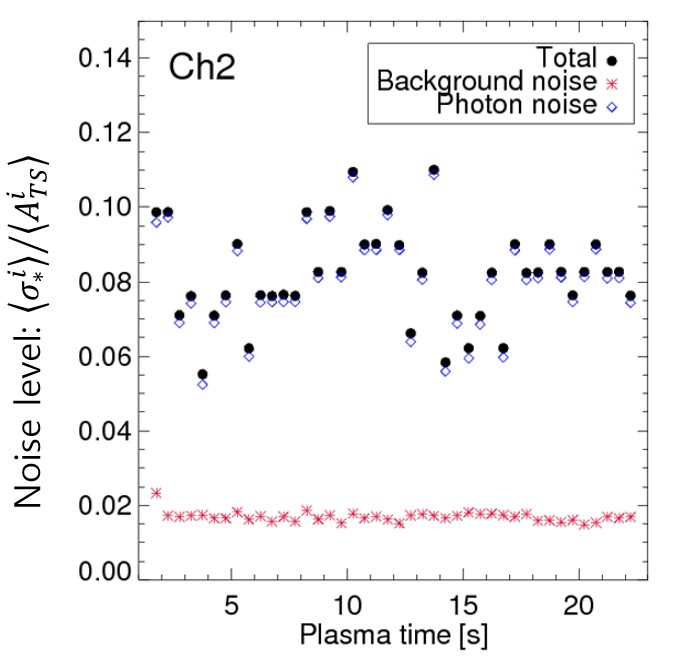}
}
\subfigure[]
{
\includegraphics[width = .43\textwidth, height = 6.5cm]{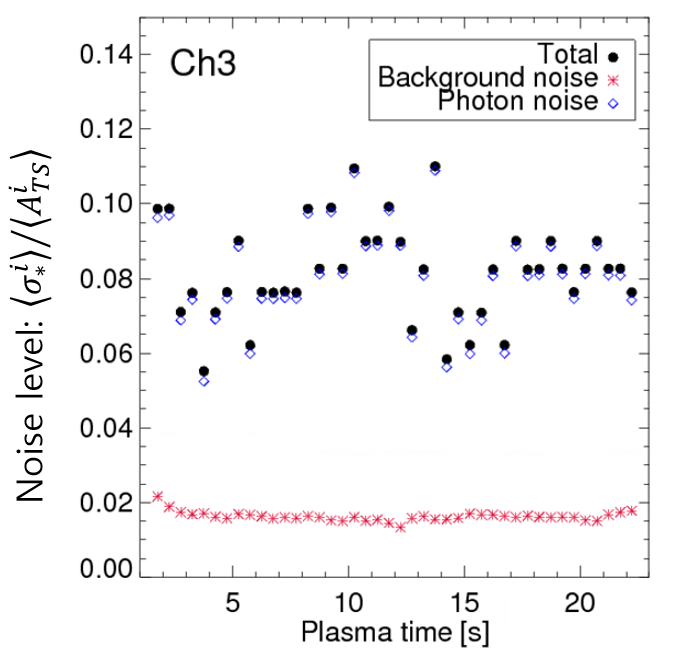}
}
\caption{Total (black circle), background (red asterisk) and photon (blue diamond) noise levels normalized to the signal levels of (a) ch 2 and (b) ch 3 for KSTAR plasma shot \#10433.}
\label{fig:proper_uncertainty}
\end{figure}

\reffig{fig:proper_uncertainty} shows, for the KSTAR \#10433, the levels of total required noise obtained by the synthetic data, background noise from experimental data and photon noise with \refeq{eq:total_noise} normalized to the experimental signal levels for chs 2 and 3. It is clear that the KSTAR Thomson scattering system during 2014 campaign is photon noise dominated at least for the chs 2 and 3 of the polychromator we have investigated. This is good that the system is not limited by the background noise, however it also suggests that the system needs to collect more photons and utilize the collected photons more efficiently to reduce the $15$\% variation on the measured $T_e$.

From the estimated normalized photon noise level, i.e., $1/SNR_{ph}^i$, we can estimate the `effective' photon counts $N_{ph}^{i,~eff}$ detected by the photon detectors as
\begin{equation}
\label{eq:photon_number}
\frac{1}{SNR_{ph}^i} = \frac{1}{\sqrt {N_{ph}^{i,~eff}}} = \frac{F}{\sqrt {N_{ph}^i}},
\end{equation}
where $F$ is the noise factor defined as the ratio of the input SNR to the output SNR \cite{Scannell_Thesis}, and $N_{ph}^i$ is the `actual' photon counts. Using the values from \reffig{fig:proper_uncertainty} and \refeq{eq:photon_number}, we find that the effective photon counts are approximately $170$ for chs 2 and 3. Note that the noise factor $F$ is always greater than or equal to $1$, thus actual photon counts may well be larger than $170$, i.e., by a factor of $F^2$. However, this cannot be regarded as the gain in photon counts since $F$ decreases the SNR.

The core KSTAR 2014 TS system has an effective F/\# of $6.7$ which corresponds to $17.5$ msr. With the incident laser wavelength of $1064$ nm and energy of $2.0$ J, the incident photon number is estimated to be $1.1\times 10^{19}$. Having approximate values of effective quantum efficiency of $15$\%, scattering length of $10$ mm, optical transmittance of $40$\%, filter transmittance of $70$\%, we estimate photon budget to be approximately 6,200 photons per $1.0\times 10^{19}$ m$^{-3}$ of electron density. For $T_e = 800$ eV, we find that the optical band-pass filters for chs 2 and 3 covers about $30$\% of the full spectrum (see \reffig{fig:KSTAR_filter_fcn}). Thus, the expected photon counts on chs 2 and 3 are approximately $1,800$ about an order of magnitude higher than the estimated effective photon counts. This means that the KSTAR TS system has large potential to be improved.

\section{Conclusion}
In this paper, we estimated the time evolution of electron temperature by using the weighted look-up table method and found a $15$\% variation on $T_e$. By using the synthetic Thomson data, we have found that such a variation on electron temperature cannot be explained solely by the background noise and thus estimated the required uncertainty level consistent with the experimental observation. Our study indicates that the chs 2 and 3 of the polychromator we have investigated are dominated by the photon noise with the average effective photon counts per laser pulse of $170$. This suggests that improving the photon collection system and decreasing the noise factor of the system is required to decrease the observed scatters in electron temperature.
\acknowledgments
This work is supported by National R\&D Program through the National Research Foundation of Korea (NRF) funded by the Ministry of Science, ICT \& Future Planning (grant number 2014M1A7A1A01029835) and the KUSTAR-KAIST Institute, KAIST, Korea.

\end{document}